\documentstyle[prb,aps,psfig]{revtex}
\begin {document}
\bibliographystyle {plain}
\twocolumn[
\hsize\textwidth\columnwidth\hsize\csname @twocolumnfalse\endcsname 
\title
{\bf Disorder and the Quantum Hall Ferromagnet.}
 \author{A. G. Green}
\address{Department of  Physics, Princeton University, New Jersey NJ 08544} 
\maketitle
\begin{abstract}
The distinguishing feature of the quantum Hall ferromagnet is the identity 
between electrical and topological charge densities of a spin distortion. In  addition to the 
wealth of physics associated with Skyrmionic excitations of the quantum Hall ferromagnet,
this identification permits a rather curious coupling of spinwaves to the disorder potential.
A wavepacket of spinwaves has an associated, oscillating dipole charge distribution,  due to the 
non-linear form of the topological density. We investigate the way in which this coupling modifies
 the conductivity and temperature dependence of magnetization of the quantum Hall ferromagnet. 
\end{abstract}
]
The distinguishing feature of the quantum Hall ferromagnet (QHF) is the identity
between the topological density of a spin  distortion and the associated electrical charge
density. This identification permits a chemical potential to stabilize topologically
non-trivial groundstate spin configurations, known as Skyrmions\cite{Sondhi}. The theoretical
 prediction of these states has received substantial experimental support\cite{Skyrmion_expts} and prompted
  a good deal of theoretical speculation. The link between topological and electrical charge 
  densities also produces a curious coupling of spinwaves to the disorder potential. Although a planewave
  spin distortion carries no charge, a wavepacket of spinwaves has an oscillating dipole charge distribution 
  associated with it, due to the non-linear form of the topological density. Spinwaves couple to the 
  disorder potential through this charge distribution.
In this work, we investigate the way in which
this coupling modifies the conductivity  and temperature dependence of 
magnetization of the quantum Hall state.

The low energy effective action for the QHF at filling
fractions $\nu=1$ and the Laughlin filling fractions is given by\cite{Sondhi,Moon}
\begin{eqnarray}
S
&=&
 \int dt d^2x 
\left[
 \frac{\bar \rho}{2} \partial_t {\bf n}.
{\bf A}[ {\bf n}]
-\frac{\rho_s}{2} |\partial_\mu {\bf n}|^2
+\bar \rho g {\bf B}. {\bf n} \right]
\nonumber\\
& & -\int dt d^2x
 J_0({\bf x}) U({\bf x})
-\int dt V \left[ J_0({\bf x}) \right]
\nonumber\\
& & +
\nu \int dt d^2x d^2y J_0({\bf x}) \epsilon_{ij} \frac{x^i-y^i}{|{\bf
x}-{\bf y}|^2} J_j({\bf y}),
\label{Effect_act}
\end{eqnarray}
where
\begin{equation}
J_{\mu}=
-\frac{e \nu}{8\pi} \epsilon_{\mu \nu \lambda} {\bf n}.
(\partial_{\nu} {\bf n} \times
\partial_{\lambda} {\bf n}).
\label{Top_charge}
\end{equation}
${\bf n} ( {\bf x})$ is an O(3)-vector order parameter of unit length, describing the 
local polarization of the quantum Hall system. The first line of Eq.(\ref{Effect_act}) is the 
usual low energy effective action for a ferromagnet.  ${\bf A}[ {\bf n}]$ is the vector potential
 of a unit monopole in spin space, $\bar \rho$ is the electron density ($\bar \rho = \nu / 2 \pi l^2$, where $l$
 is the magnetic length), $\rho_s$ is the spin stiffness and $g$ is the Zeeman coupling, into which we
 have absorbed the electron spin and the Bohr magneton for ease of notation. The second line 
 of Eq.(\ref{Effect_act}) contains terms arising due to the identity of charge and topological charge
 (which is embodied in Eq.(\ref{Top_charge})). The first of these terms is an interaction with the disorder 
 potential, $U({\bf x})$, and the second, $V[J_0({\bf x})]$, is the Coulomb energy of the charge distribution,
 $J_0({\bf x})$. Eq.(\ref{Effect_act}) describes both the low energy spin and charge dynamics of the quantum 
 Hall system. The quantization of Hall conductivity follows from the final term, the Hopf term\cite{Yakovenko}.
 
Here, we are concerned with the effect of the disorder
potential upon small fluctuations, ${\bf l}=(l_1,l_2,0)$, about the
ferromagnetic groundstate, $\bar {\bf
n}=(0,0,1)$;  ${\bf n}=(l_1,l_2,\sqrt{1-|{\bf l}|^2})$. The effective
action and current, expanded to lowest order in these fluctuations, are
\begin{eqnarray}
S&=&
\int d^2x dt \frac{1}{2}  \bar l \left( \frac{\bar \rho}{2}
\partial_t - \rho_s \nabla^2- \bar \rho g B \right)  l
\nonumber\\
& &
-\int d^2x dt J_0({\bf x}) U({\bf x}),
\nonumber\\
J_{\mu}&=&
i\frac{e \nu}{8 \pi} \epsilon^{\mu \nu \lambda}
\partial_{\nu} \bar l \partial_{\lambda} l.
\label{complex_l_action}
\end{eqnarray}
We use the complex notation, $l=l_1+il_2$, $\bar l=l_1-il_2$.
Both the Coulomb and statistical interactions have been neglected in writing
down Eq.(\ref{complex_l_action}). Although important in determining the size and
shape of the Skyrmion excitations, the former is less relevant than the 
remaining terms in its effect upon spinwaves\cite{Read}. We will show later that the quantization of Hall 
conductivity, produced by the Hopf term, is unaffected by weak disorder.
The calculations presented in this work concern the perturbative effects of weak
disorder. It is worth noting that the effective action, Eq.(\ref{complex_l_action}), is very
similar to that of electrons in a random potential, aside from the unusual form of the
current density and the bosonic nature of the fields. This similarity is suggestive of the 
possibility of weak localization  effects. These are not considered here.

We represent the bare, momentum space propagators, $\langle \bar l ({\bf q},
\tilde \omega) l(-{\bf q}, -\tilde \omega) \rangle$ and $\langle
 \partial_{\mu} \bar l ({\bf q}, \tilde \omega) \partial_{\nu}
l(-{\bf q}, -\tilde \omega)  \rangle$, by the diagrams
\begin{eqnarray*}
\langle \bar l ({\bf q},
\tilde \omega) l(-{\bf q}, -\tilde \omega) \rangle
&=&
\begin{picture}(47,12)(0,0)
\psfig{file=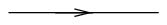}
\end{picture}
=\frac{1}{i\bar \rho \tilde \omega/2 - E({\bf q})}
\\
\langle \partial_{\mu} \bar l
({\bf q}, \tilde \omega) \partial_{\nu} l (-{\bf q}, -\tilde \omega) \rangle
&=&
\begin{picture}(47,12)(0,3)
\psfig{file=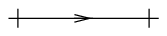}
\end{picture}
=\frac{q_{\mu} q_{\nu}}{i\bar \rho \tilde \omega/2 - E({\bf q})}
\end{eqnarray*}
where $E({\bf q})=\rho_s |{\bf q}|^2 + \bar \rho g B$ is the spin energy density.
The disorder interaction is given by
$$
S_i =  \! \int \! d \tilde \omega \frac{d^2q_1}{(2\pi)^2}  \frac{d^2q_2}{(2\pi)^2}
i \left( \frac{e \nu}{8 \pi} \right) \epsilon^{ij} 
\left(
\begin{picture}(75,25)(2,15)
\psfig{file=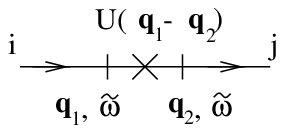}
\end{picture}
\right),
$$
where the frequency integral, $\int d \tilde \omega$, is a shorthand notation for
the bosonic Matsubara frequency summation $\frac{1}{T}
\sum_{n=-\infty}^{\infty}... |_{\tilde \omega = 2\pi n/T}$. 
Notice that the scattering off the impurity potential is entirely
elastic {\it i.e.} the energy labels on the propagators are
conserved.

In GaAs heterostructures, the disorder potential felt by the electrons
in the 2DEG is due mainly to Coulomb interaction with ionized donor
impurities in the n-type region\cite{Efros1}. This region is separated
from the 2DEG by 
an insulating spacer layer of width $d$. 
One may obtain an expression for the
correlations in the disorder potential
by modeling this situation
with the potential due to a random planar distribution of charge at a
distance $d$  from the 2DEG. 
The correlations in the disorder potential in this model are given by
\begin{eqnarray}
\langle \langle U_{\bf q} U_{{\bf q}'} \rangle \rangle
&=&
(2\pi)^2 \delta({\bf q} + {\bf q}') 
\left(\frac{e \sqrt{n_d}}{2\epsilon} \right)^2
\frac{e^{-2|{\bf q}|d}}{|{\bf q}|^2}
\nonumber\\
&=&
(2\pi)^2 \delta({\bf q} + {\bf q}') 
\left(
\begin{picture}(48,10)(0,5) \psfig{file=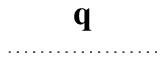}
\end{picture} 
\right),
\label{UU}
\end{eqnarray}
where $n_d$ is the area density of donor impurities. This simple model
of disorder somewhat overestimates the potential felt by the
2DEG. Due to Coulomb interactions between the donors, the size of the
fluctuations in the disorder potential is usually much less than would
be expected for a totally uncorrelated distribution of charge in the
disorder plane. We follow Fogler {\it et al.}\cite{Fogler} and assume that
this  effect may be taken into account by interpreting $n_d$ in
Eq.(\ref{UU}) as a density of `uncorrelated' donors, which is much
less than the actual density of donors. 

The lowest order contribution of disorder to the self-energy is
\begin{eqnarray}
\Sigma (i\tilde \omega, {\bf p})
&=&
\begin{picture}(120,30)(-10,0)
\psfig{file=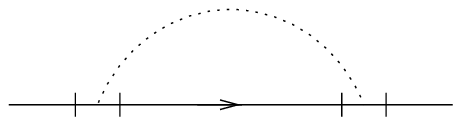}
\end{picture}
\nonumber\\
&=&
K \rho_s^2 
\int \frac{d^2q}{(2\pi)^2}
\frac{ ({\bf p} \times {\bf q})^2}{i \bar \rho \tilde \omega /2 -
E({\bf q}+{\bf p})} 
\frac{e^{-2 d |{\bf q}|}}{|{\bf q}|^2},
\end{eqnarray}
where 
\begin{equation}
K=
\frac{1}{\rho_s^2}	
\left( \frac{e \nu}{8 \pi} \right)^2
\left( \frac{e \sqrt{n_d}}{2 \epsilon} \right)^2
\label{K}
\end{equation}
is a dimensionless measure of the disorder strength.
The retarded self-energy is obtained by analytic
continuation to real frequencies with the
substitution $i \tilde \omega \rightarrow \omega + i \delta$. The real
and imaginary parts of the self-energy so obtained are
\begin{eqnarray}
{\cal R}e \Sigma(\omega, {\bf p})
&=&
K \rho_s^2
\int \frac{d^2q}{(2\pi)^2}
\frac{ ({\bf p} \times {\bf q})^2}{\bar \rho \omega /2 -
E({\bf q}+{\bf p})} 
\frac{e^{-2 d |{\bf q}|}}{|{\bf q}|^2},
\label{real_part}
\end{eqnarray}
\begin{eqnarray}
{\cal I}m \Sigma(\omega, {\bf p})
&=&
 K \rho_s^2
\int \frac{d^2q}{(2\pi)^2}
({\bf p} \times {\bf q})^2
\frac{ e^{-2 d |{\bf q}|}}{|{\bf q}|^2}
\nonumber\\
& &\;\;\;\;\;\;\;\;\;\;\;\;\;\;\;\;
\times
\pi \delta( \bar \rho \omega /2 - E({\bf q}+{\bf p}) ).
\label{imaginary_part}
\end{eqnarray}
The real part of the self-energy can be approximated from Eq.(\ref{real_part}) in the 
limit $\rho_s|{\bf p}|^2$, $\bar \rho| \omega-2gB|/2 \ll \rho_s/d^2$. The leading order
contribution is proportional to $|{\bf p}|^2$ and provides a correction to the spinwave
stiffness, $\Delta \rho_s= {\cal R}e \Sigma/|{\bf p}|^2$. For $\rho_s |{\bf p}|^2 > \bar \rho| \omega-2gB|/2$,
there is a crossover to $|{\bf p}|^2 \ln |{\bf p}|^2$ dependence. We find
\begin{eqnarray}
{\cal R}e \Sigma(\omega, {\bf p})
&\simeq&
\frac{K \rho_s}{8 \pi}
|{\bf p}|^2 
\ln \left[ \frac{4|\bar \rho g B -\bar \rho \omega/2|d^2}{\rho_s}
\right]
\nonumber\\
& &
\hbox{ for } \rho_s |{\bf p}|^2 < \bar \rho| \omega-2gB|/2,
\nonumber\\
&\simeq&
\frac{K \rho_s}{8 \pi}
|{\bf p}|^2 
\ln \left[ 4 |{\bf p}|^2 d^2 \right]
\nonumber\\
& &
\hbox{ for } \rho_s |{\bf p}|^2 > \bar \rho| \omega-2gB|/2.
\label{real_sigma}
\end{eqnarray}
The first of these expressions has been calculated by expanding 
Eq.(\ref{real_part}) to lowest order in $|{\bf p}|^2$ and by replacing the exponential 
factor, $e^{-2 d |{\bf q}|}$, with an ultra-violet cut-off, $1/2d$. The second expression 
is calculated exactly from Eq.(\ref{real_part}), setting $\omega=2gB$.

The imaginary part of the self energy may be calculated exactly
when $d=0$, with the result
\begin{eqnarray}
{\cal I}m \Sigma(\omega, {\bf p})
&=&
-
\frac{K }{8}
\bar \rho(\omega/2 -gB)
\theta \left( \omega/2 -gB \right)
\nonumber\\
& & \hbox{for  } \rho_s |{\bf p}|^2 > \bar \rho| \omega-2gB|/2,
\nonumber\\
&=&
-
\frac{K }{8}
\rho_s |{\bf p}|^2
\nonumber\\
& & \hbox{for  } \rho_s |{\bf p}|^2 < \bar \rho| \omega-2gB|/2.
\label{Im_self_energy}
\end{eqnarray}
The integral for finite $d$ is much trickier and cannot be carried out
analytically. For large $d$ it is exponentially suppressed by
a factor $e^{-2d|{\bf p}|}$.

Taken at face value,
Eq.(\ref{real_sigma}) implies a threshold disorder strength at which the renormalized
spin-stiffness is zero at zero frequency.
We interpret this as indicative of a depolarization transition to a paramagnetic state. A similar
suggestion has been made by Fogler {\it et al.}\cite{Fogler} in order to explain the breakdown 
of spin splitting in high Landau levels. Strictly, the calculations presented here apply only for 
weak disorder and small $\Delta \rho_s$. That the threshold behaviour suggested here does indeed
 occur, may be seen in a number of ways. The most elegant of these is through a Bogomolny bound 
 type argument\cite{Cooper}. The present treatment enables one to investigate
  the approach to this threshold.

{\bf Optical conductivity.}
The longitudinal and transverse conductivities are given
by the Kubo formula\cite{Greensfunctions}:
\begin{eqnarray}
\sigma_{ij}(\omega)
&=&
\left.
\frac{i}{\omega} 
\langle J_i(0,\tilde \omega) J_j(0,-\tilde \omega) \rangle
\right|_{i\tilde \omega \rightarrow \omega +i \delta}
\label{Kubo}
\end{eqnarray}
In order to determine the longitudinal conductivity, we must evaluate
the following diagram:
\begin{eqnarray}
\lefteqn{ \langle {\bf J}(0,\tilde \omega).
 {\bf J}(0,-\tilde \omega) \rangle}
 \nonumber\\
&=&
- \left( \frac{e \nu}{8 \pi} \right)^2
\epsilon^{i \alpha \beta} \epsilon^{i \gamma \delta}
\raisebox{-7ex}{\em
\begin{picture}(150,60)(0,0)
\psfig{file=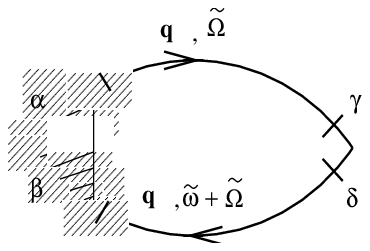}
\end{picture}}
\nonumber\\
&=& - \left( \frac{e \nu}{8 \pi} \right)^2
\epsilon^{i \alpha \beta} \epsilon^{i \gamma \delta}
\int \frac{d^2q}{(2 \pi)^2} d \tilde \Omega
\;\;  q_{\mu} \tilde q_{\nu}  q_{\gamma} \tilde q_{\delta}
\nonumber\\
& & \times
\Gamma_{\alpha \beta, \mu \nu} ({\bf q}, i\tilde \Omega, i\tilde \Omega + i\tilde \omega)
{\cal G}({\bf q}, i\tilde \Omega + i\tilde \omega)
{\cal G}({\bf q}, i\tilde \Omega),
\label{current_correlator}
\end{eqnarray}
where $q_{\mu} = (i\tilde \Omega,{\bf q})$, $\tilde q_{\mu} = (i\tilde
\Omega  + i\tilde \omega ,{\bf q})$ and ${\cal G}({\bf q}, i\tilde
\Omega)$ is the full thermodynamic Green's function. The vertex function, $
\Gamma_{\alpha \beta, \mu \nu} $, is given by the summation
\begin{eqnarray}
\lefteqn{
\Gamma_{\alpha \beta, \mu \nu} ({\bf q}, i\tilde \Omega, i\tilde
\Omega + i\tilde \omega) }
\nonumber\\
&=&
\delta_{\alpha \mu} \delta_{\beta \nu}
+
\left( \frac{e \nu}{8 \pi} \right)^2
\epsilon^{\alpha' \mu} \epsilon^{\beta' \nu}
\raisebox{-5ex}{\em
\begin{picture}(100,25)(0,0)
\psfig{file=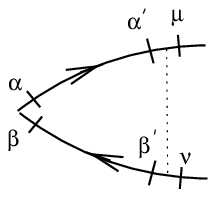}
\end{picture}}
\nonumber\\
& &
+...
\raisebox{-5ex}{\em
\begin{picture}(75,50)(-5,0)
\psfig{file=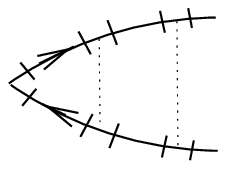}
\end{picture}}
+...
\raisebox{-5ex}{\em
\begin{picture}(75,50)(-5,0)
\psfig{file=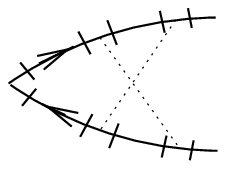}
\end{picture}}
+...
\end{eqnarray}
In fact, all contributions to the vertex function contain a factor of
$  q_{\alpha} \tilde q_{\beta}$ and there is considerable
simplification in defining a new, scalar vertex function, $\gamma({\bf
q}, i\tilde \Omega, i\tilde \Omega + i\tilde \omega)$;
$$
 q_{\alpha} \tilde q_{\beta}
\gamma({\bf q}, i\tilde \Omega, i\tilde \Omega + i\tilde \omega)
=
\Gamma_{\alpha \beta, \mu \nu} ({\bf q}, i\tilde \Omega, i\tilde
\Omega + i\tilde \omega)
 q_{\mu} \tilde q_{\nu}.
$$
This definition of the vertex function is then  substituted into Eqs.(\ref{Kubo},\ref{current_correlator})
to find the conductivity. After performing the summation over bosonic Matsubara frequencies and 
a few other standard manipulations\cite{Greensfunctions}, the real part of the longitudinal conductivity
is given by the expression
\begin{eqnarray}
\lefteqn{
\sigma (\omega)
=
\omega
\left( \frac{e \nu}{8 \pi} \right)^2
\int \frac{d^2q}{(2 \pi)^2}
|{\bf q}|^2
\int_{-\infty}^{\infty} \frac{d \epsilon}{4 \pi}
\left[ n_B(\epsilon+\omega) - n_B(\epsilon) \right]
}
\nonumber\\
& &
\times
\Re e
\left[
G^A({\bf q}, \epsilon) 
G^R({\bf q},\epsilon+\omega)
\gamma({\bf q},\epsilon - i \delta, \epsilon + \omega + i \delta)
\right.
\nonumber\\
& &
\left.
\;\;\;\;\;\;\;\;\;
-
G^R({\bf q}, \epsilon) 
G^R({\bf q},\epsilon+\omega)
\gamma({\bf q},\epsilon + i \delta, \epsilon + \omega + i \delta) 
\right],
\nonumber\\
\label{general_conductivity}
\end{eqnarray}
where $n_B(x)$ is the Bose occupation number. The contribution to the 
Hall conductivity is zero, on symmetry grounds,  since the current-current correlator $\langle {\bf J}
\times {\bf J} \rangle$  gives rise to a factor of ${\bf q} \times {\bf q} $ in the integrand. Compared
with the analogous result for electronic conductivity\cite{Greensfunctions}, Eq.(\ref{general_conductivity})
contains an additional factor of $\omega^2$, which ensures that the d.c. conductivity is zero. This is due to 
the fact that the charge fluctuations in the QHF are dipolar.

{\bf Vertex corrections.}
In the ladder approximation, the vertex function is given by the
following Dyson's equation:
\begin{eqnarray*}
\lefteqn{
\Gamma_{\alpha \beta, \mu \nu}
({\bf q}, i \tilde \Omega, i \tilde \omega  + i \tilde \Omega)
=
\delta_{\alpha \mu} \delta_{\beta \nu}
}
\nonumber\\
& &
-
\left( \frac{e \nu}{8 \pi} \right)^2
\left( \frac{e^2 n_d}{2 \epsilon} \right)^2
\int \frac{d^2k}{(2 \pi)^2}
\epsilon^{b \nu} \epsilon^{d \mu}
 k_a k_b k_c k_d
\frac{e^{-2d|{\bf q}-{\bf k}|}}{|{\bf q}-{\bf k}|^2}
\nonumber\\
& &
\;\;\;\;\;\;\;
\times
{\cal G}({\bf k}, i\tilde \Omega + i\tilde \omega)
{\cal G}({\bf k}, i\tilde \Omega)
\Gamma_{\alpha \beta, ac}
({\bf k}, i \tilde \Omega, i \tilde \omega  + i \tilde \Omega),
\end{eqnarray*}
or diagrammatically,
$$
\begin{picture}(250,65)(0,10)
\psfig{file=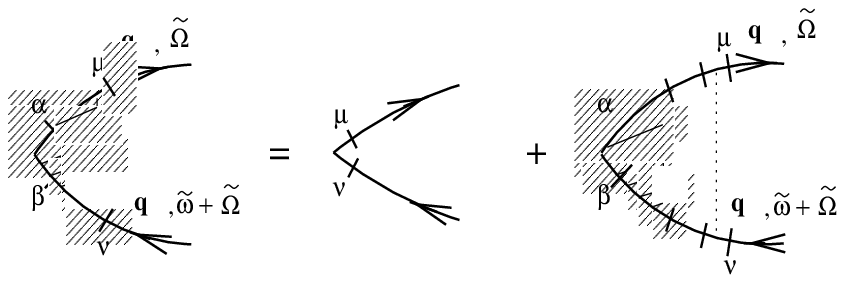}
\end{picture}.
$$
This Dyson's equation may be recast in terms of the scalar vertex
function, $\gamma ({\bf q}, i \tilde \Omega, i \tilde \omega  + i
\tilde \Omega)$:
\begin{eqnarray}
\lefteqn{
\gamma({\bf q}, i \tilde \Omega, i \tilde \omega  + i
\tilde \Omega)
=
1
+
\int \frac{d^2k}{(2 \pi)^2}
\frac{
\left( {\bf q} . {\bf k} \right)^2}
{|{\bf q}|^4}
W_{ {\bf q}  {\bf k}}
}
\nonumber\\ & &
\;\;\;
\times
{\cal G}({\bf k}, i\tilde \Omega + i\tilde \omega)
{\cal G}({\bf k}, i\tilde \Omega)
\gamma
({\bf k}, i \tilde \Omega, i \tilde \omega  + i \tilde \Omega),
\label{gamma_Dyson}
\end{eqnarray}
where
$$
W_{ {\bf q}  {\bf k}}=
-
\left( \frac{e \nu}{8 \pi} \right)^2
\left( \frac{e^2 n_d}{2 \epsilon} \right)^2
\left( {\bf q} \times {\bf k} \right)^2 
\frac{e^{-2d|{\bf q}-{\bf k}|}}{|{\bf q}-{\bf k}|^2}.
$$
In order to obtain Eq.(\ref{gamma_Dyson}), we have 
used the relation
$$
\Gamma_{ab, cd}
({\bf k}, i \tilde \Omega, i \tilde \omega  + i \tilde \Omega)
  q_a q_b k_c k_d
= 
\gamma ({\bf k}, i \tilde \Omega, i \tilde \omega  + i \tilde \Omega)
\left( {\bf k}. {\bf q} \right)^2,
$$
which follows from the definition of $\gamma$ and the symmetry of the 
disorder interaction; since the disorder potential couples only to the 
charge density and not to any other components of the current density,
$\Gamma_{\alpha \beta, \mu \nu}= \delta_{\alpha \mu} \delta_{\beta \nu}$
 if either or both of $\mu$ or $\nu$ are time-like.

In order to calculate the real part of the optical conductivity we require
$\gamma({\bf q}, \epsilon-i\delta, \epsilon+\omega+i\delta)$. Evaluating this is a very 
difficult task. However, several simplifying assumptions may be made. Firstly,
we assume that the frequency dependence of the optical conductivity is dominated
by terms in Eq.(\ref{general_conductivity}) other that the vertex function. Secondly,
the terms $G^AG^R$ in Eq.(\ref{general_conductivity}) are strongly peaked within $\omega$ of
$\epsilon=2E({\bf q})/\bar \rho$. Therefore, we have only to calculate
$\gamma({\bf q},2E({\bf q})/\bar \rho) =\gamma({\bf q},2E({\bf q})/\bar \rho-i\delta,2E({\bf q})/\bar \rho-i\delta)$.
Using Eq.(\ref{gamma_Dyson}), we find
\begin{eqnarray}
\lefteqn{\gamma({\bf q}, 2E({\bf q})/\bar \rho)}
\nonumber\\
& &
=
1
+
\left.
\int \frac{d^2k}{(2 \pi)^2}
\frac{
\left( {\bf q} . {\bf k} \right)^2}
{|{\bf q}|^4}
W_{ {\bf q}  {\bf k}}
G^R({\bf k},\epsilon) G^A({\bf k},\epsilon)
\gamma({\bf k}, \epsilon) 
\right|_{\epsilon=2E({\bf q})/\bar \rho}
\nonumber\\
& &
=
1
+
\left.
\int \frac{d^2k}{(2 \pi)^2}
\frac{
\left( {\bf q} . {\bf k} \right)^2}
{|{\bf q}|^4}
W_{ {\bf q}  {\bf k}}
\frac{A({\bf k}, \epsilon)}{2 \Delta ({\bf k}, \epsilon)},
\gamma({\bf k}, \epsilon)
\right|_{\epsilon=2E({\bf q})/\bar \rho},
\label{gamma_Dyson_simple}
\end{eqnarray}
where $\Delta({\bf k}, \epsilon)=-{\cal I}m \Sigma ({\bf k}, \epsilon)$ and 
$A({\bf k}, \epsilon)=-2{\cal I}m G^R({\bf k}, \epsilon)$ is the spectral function.
In the limit of very weak disorder, $A({\bf k}, \epsilon)\approx 2\pi \delta(\bar \rho \epsilon/2-E({\bf k}))$.
The delta function imposes the constraint $|{\bf k}|=|{\bf q}|$ and, since $\gamma({\bf q}, 2E({\bf q})/\bar \rho) \equiv
\gamma(|{\bf q}|)$, Eq.(\ref{gamma_Dyson_simple}) reduces to an algebraic equation. 
The solution is
\begin{eqnarray}
\gamma({\bf q}, 2E({\bf q})/\bar \rho) &=& \frac{ \Delta({\bf k},2E({\bf q})/\bar \rho) }{\Delta_T({\bf k},2E({\bf q})/\bar \rho) }
\\
 \Delta({\bf k},2E({\bf q})/\bar \rho) &=&
\int \frac{d^2k}{(2 \pi)^2} W_{ {\bf q}  {\bf k}} \delta( E({\bf q})-E({\bf k}) )
\label{self_energy_2}
\\
\Delta_T({\bf k},2E({\bf q})/\bar \rho) &=&
\int \frac{d^2k}{(2 \pi)^2} W_{ {\bf q}  {\bf k}} \delta( E({\bf q})-E({\bf k}) )
\nonumber\\
& &\;\;\;\;\;\;\;\;\;\;\;\;\;\;\;\;\;\;
\times
\left( 1- 
\frac{
\left( {\bf q} . {\bf k} \right)^2}
{|{\bf q}|^4}
\right)
\label{deltaT}
\end{eqnarray}
Eq.(\ref{self_energy_2}) is simply a re-writing of Eq.(\ref{imaginary_part}) for the 
imaginary part of the spinwave self-energy. The final term in the integrand of Eq.(\ref{deltaT}) 
is an angular weighting, $\sin^2 \theta$, for scattering events, where $\theta$ is the angle 
between incoming and outgoing spinwave states. This should be compared with the
electronic case, where the angular weighting is $1-\cos \theta$.

{\bf Ignoring vertex corrections}
(substituting $\gamma=1$),  Eq.(\ref{general_conductivity}) reduces to
\begin{eqnarray}
\sigma (\omega)
&=&
\omega
\left( \frac{e \nu}{8 \pi} \right)^2
\int \frac{d^2q}{(2 \pi)^2}
|{\bf q}|^2
\int_{-\infty}^{\infty} \frac{d \epsilon}{4 \pi}
\nonumber\\
& & \times
\left[ n_B(\epsilon+\omega) - n_B(\epsilon) \right]
A({\bf  q},\epsilon) A({\bf q}, \epsilon+\omega).
\label{sigma_without_gamma}
\end{eqnarray}
A similar calculation of the finite wavevector conductivity, neglecting vertex corrections,
gives
\begin{eqnarray*}
\sigma (\omega,{\bf k})
&=&
\frac{1}{\omega}
\left( \frac{e \nu}{8 \pi} \right)^2
\int \frac{d^2q}{(2 \pi)^2}
\int_{-\infty}^{\infty} \frac{d \epsilon}{4 \pi}
|\omega {\bf k}- \epsilon {\bf q}|^2
\nonumber\\
& & \times
\left[ n_B(\epsilon+\omega) - n_B(\epsilon) \right]
A({\bf  q},\epsilon) A({\bf q}+{\bf k}, \epsilon+\omega).
\end{eqnarray*}
In contrast to the zero wavevector conductivity, $\sigma (\omega,{\bf k})$ may be non-zero
in the absence of disorder.
Eq.(\ref{sigma_without_gamma}) may now be used, in conjuction with the spinwave self-energy,
Eqs.(\ref{real_sigma},\ref{Im_self_energy}), in order to calculate the contribution of disorder scattered
spinwaves to the optical conductivity. In the absence of disorder, the spectral function has a 
single delta-function peak, $A({\bf q},\epsilon)=2 \pi \delta(\bar \rho \epsilon/2 - E({\bf q}))$. The
effect of disorder is to broaden and shift this peak.
For $T \ll g, \omega$ and weak disorder, the product $A({\bf  q},\epsilon) A({\bf q}, \epsilon+\omega)$,
derived from Eqs.(\ref{real_sigma},\ref{Im_self_energy}), is strongly peaked at 
$\bar \rho \epsilon/2 = E({\bf q})$ and $\bar \rho ( \epsilon+\omega)/2 = E({\bf q})$
and may be approximated by
\begin{eqnarray*}
A({\bf  q},\epsilon) A({\bf q}, \epsilon+\omega)
&\approx&
2 \pi \delta 
\left( \bar \rho \epsilon/2 - E({\bf q}) \right)  A({\bf q}, \epsilon+\omega)
\\
&+&
2 \pi \delta
\left( \bar \rho (\epsilon+\omega)/2 - E({\bf q}) \right) A({\bf q}, \epsilon)
.
\end{eqnarray*}
The real part of the longitudinal optical conductivity, calculated within this approximation, is
\begin{eqnarray}
\sigma(\omega)
&\approx&
\frac{K}{32 \pi \rho_s^2}
\left( \frac{e \nu}{8 \pi} \right)^2
T^2 (1-e^{-\omega/T}) e^{-2gB/T}.
\label{cond1}
\end{eqnarray}
At very small frequency, $\omega \ll KT$, the product $A({\bf  q},\epsilon) A({\bf q}, \epsilon+\omega)$
is no longer resolved into two peaks. The dominant frequency dependence in Eq.(\ref{sigma_without_gamma})
then comes from the $n_B(\epsilon)-n_B(\epsilon+\omega)$ term. Then 
$$A({\bf  q},\epsilon) A({\bf q}, \epsilon+\omega) \approx A^2({\bf q}, \epsilon)
=
\frac{2 \pi \delta (\bar \rho \epsilon/2- E({\bf q}))}{ \Im m \Sigma({\bf q},\epsilon)}
.
$$
The energy and momentum integrals in Eq.(\ref{sigma_without_gamma}) may then be carried out with the result
\begin{eqnarray}
\sigma(\omega)
& \approx&
\frac{1}{\pi \rho_s^2 K}  \left( \frac{e \nu}{8 \pi} \right)^2
\omega^2 e^{-2gB/T}
\hbox{ for } g \gg T.
\label{cond2}
\end{eqnarray}
For typical experimental systems at $\nu=1$, an upper estimate for the disorder strength
is $K \sim 0.1$ (approximating $n_d=\bar \rho$) and the spin stiffness $\rho_s\sim 4K$. 
The conductivities predicted by Eqs.(\ref{cond1},\ref{cond2}) are vanishingly small
and probably unmeasurable.

{\bf Magnetization}
The variation of magnetization with temperature, in the absence of disorder, has 
been calculated by Read and Sachdev\cite{Read}, using a lowest order $1/N$ 
expansion. We extend this calculation to include the effect of disorder. 
Firstly, a Hopf map (${\bf n}=\bar z_{\alpha} {\bf \sigma}_{\alpha \beta}
z_{\beta}$, $\sum_{\alpha=1}^{2} |z_\alpha|^2=1$) is used to recast
the effective action, Eq.(\ref{Effect_act}), into CP1 form;
\begin{eqnarray}
S&=&
\int d^2x dt 
\left[
i\frac{\bar \rho}{2} \bar z \partial_t z
+\rho_s |D_i z|^2
+\bar \rho g B \bar z \sigma^z z
\right]
\nonumber\\
&-&
\int d^2x dt 
\left[
U({\bf x}) J_0({\bf x})
+\lambda \left( |z|^2-1 \right)
\right],
\nonumber\\
J_{\nu}
&=&
-\frac{i \nu e}{2 \pi} \epsilon^{\mu \nu \lambda}
\partial_{\nu} \bar z_{\alpha} \partial_{\lambda} z_{\alpha}
\label{CP1}
\end{eqnarray}
where $D_i = \partial_i+i \theta_i $. $\theta_i $ is an auxiliary
field, introduced in order to decouple quartic terms in the effective
action. $\lambda$ is a Lagrange multiplier that imposes the
constraint. The indices on $z_{\alpha}$ have been suppressed for clarity.

To zeroth order in the $1/N$ expansion, the constraint is imposed at the mean field level in
order to self-consistently determine the average value of the Lagrange
multiplier, $\bar \lambda$\cite{comment}. The resulting gap equation is
\begin{eqnarray}
\langle \langle \bar z  z \rangle \rangle
&=&
\sum_{\sigma= \pm} \int \frac{d^2p}{(2\pi)^2} d\tilde \Omega
\bar {\cal G}(i \tilde \Omega, {\bf p}^2, \sigma,\bar \lambda)
\nonumber\\
&=&
\sum_{\sigma= \pm} \int \frac{d^2p}{(2\pi)^2} \int_{-\infty}^{\infty} 
\frac{d\epsilon}{2\pi} n_B(\epsilon) 
A(\epsilon, {\bf p}^2, \sigma, \bar \lambda),
\label{gap_equation}
\end{eqnarray}
where $\bar {\cal G}( i\tilde \Omega, {\bf p}^2, \sigma, \bar \lambda)$
indicates the disorder average of the $\bar z z $-Green's
function and 
$A(\epsilon, {\bf p}^2, \sigma, \bar \lambda)=-2 \Im m 
\bar G_{ret}( \epsilon, {\bf p}^2, \sigma,\bar \lambda)$ 
is the spectral function. We have carried out the frequency
summation in order to obtain the final expression.
The magnetization may also be calculated to this order and is given by
\begin{eqnarray}
\langle \langle \bar z \sigma^z z \rangle \rangle
&=&
\sum_{\sigma= \pm} \int \frac{d^2p}{(2\pi)^2} d\tilde \Omega
\sigma \bar {\cal G}( i\tilde \Omega, {\bf p}^2, \sigma,\bar \lambda)
\nonumber\\
&=&
\sum_{\sigma= \pm} \sigma \int \frac{d^2p}{(2\pi)^2} \int_{-\infty}^{\infty}  
\frac{d\epsilon}{2\pi} n_B(\epsilon) 
A(\epsilon, {\bf p}^2, \sigma, \bar \lambda).
\label{magnetization}
\end{eqnarray}
To O(1/N), Eq.(\ref{CP1}), is identical to the sum of 
two copies of the spinwave action, Eq.(\ref{complex_l_action}), with the Zeeman term, $\bar \rho gB$, replaced with
$\sigma \bar \rho gB+\bar \lambda$. The expressions for the self-energy derived above may be used directly 
with this replacement. In the absence of disorder, the spectral function has a single delta-function peak;
$A(\epsilon, {\bf p}^2, \sigma, \bar \lambda) = 
2\pi \delta (\bar \rho \epsilon/2 - E ({\bf p}^2, \sigma, \bar \lambda)$), where  
$ E ({\bf p}^2,\sigma, \bar \lambda)
=  \rho_s {\bf p}^2 +\sigma \bar \rho gB+\bar \lambda$. 
Substitution of this into Eqs.(\ref{gap_equation},\ref{magnetization}), reproduces the result of 
[\onlinecite{Read}].
The effect of disorder is to broaden and shift this peak. 
The real part of the self-energy produces a renormalization of the spin stiffness, $\rho_s \rightarrow \tilde \rho_s$.
Upon direct substitution of Eq.(\ref{Im_self_energy}), one finds that, to lowest order in $K$, the new position 
of the peak is at
$
\bar \rho \epsilon/2
=
\tilde E
-4K^2 \tilde \rho_s {\bf p}^2
$
and so the shift due to the imaginary part of the self-energy may be incorporated as a further renormalization of the spin-stiffness.
This is the dominant effect of weak disorder. The gap equation and magnetization
are given by the disorder free expressions \cite{Read}
with appropriately renormalized spin-stiffness\cite{ON_expansion}.

The calculation of Ref.[\onlinecite{Read}] shows good agreement with experiment\cite{Magnetization}
 aside from at high temperatures, where the experimentally measured magnetization appears to
 fall below even the theoretical $\rho_s=0$ prediction. Recent work\cite{Recent} has 
 shown that this discrepancy cannot be explained by the inclusion of higher orders in the 1/N expansion.
 Here, we have shown that neither can it be explained by the effects of weak disorder. 
 In fact, to explain this observation would require spectral weight to be transfered below the Zeeman gap.
 This appears to be impossible so long as the groundstate remains ferromagnetic. Two possible 
 alternative explanations lie in the effect of Skyrmions or the inclusion of the correct spinwave
 dispersion at high momenta.
 The latter approach has provided a good explanation for the dramatic reduction in magnetization with increasing
 temperature found at $\nu=1/3$\cite{Turberfield}. It is readily incorporated into the lowest order 1/N expansion in the
 absence of disorder, by inserting a spectral function with a delta-function peak at the correct spinwave dispersion
 into Eq.(\ref{gap_equation},\ref{magnetization}) and solving the resulting equations numerically.
 
In conclusion, we have considered the effect of weak disorder upon the quantum Hall ferromagnet. The identification
of charge and topological charge of spinwave distortions allows a coupling of spins to the disorder potential. The 
signature of this coupling in the temperature dependence of magnetization is a reduction of the effective spin-stiffness.
The effect upon conductivity is rather more interesting, although unfortunately it is probably unmeasurably small.
 We predict a spinwave contribution to the longitudinal optical
conductivity at finite temperature.

We acknowledge A. M. Tsvelik, J. T. Chalker and S. Sondhi for helpful comments and suggestions.


\begin{thebibliography}{99}
\bibitem{ON_expansion} A similar analysis may be carried out for the 1/N expansion in the O(3) representation\cite{Read}.
The appropriate gap equation and magnetization are modified in the same way as in the CP1 representation.
\end{thebibliography}
\end{document}